\documentclass[%
 reprint,
preprintnumbers,
 amsmath,amssymb,
 aps,
]{revtex4-2}

\usepackage{graphicx}
\usepackage{dcolumn}
\usepackage{bm}
\usepackage{hyperref}
\hypersetup{colorlinks=true, citecolor=blue, urlcolor=blue, linkcolor=blue}

\usepackage[alsoload=hep]{siunitx} 

\usepackage{xcolor} 

\usepackage[caption=false]{subfig}

\newcommand\snowmass{\begin{center}\rule[-0.2in]{\hsize}{0.01in}\\\rule{\hsize}{0.01in}\\
\vskip 0.1in Submitted to the  Proceedings of the US Community Study\\ 
on the Future of Particle Physics (Snowmass 2021)\\ 
\rule{\hsize}{0.01in}\\\rule[+0.2in]{\hsize}{0.01in} \end{center}}

\begin{document}
\title{DIMUS: Super-Compact Dimuonium Spectroscopy Collider at Fermilab}
\author{Patrick J.~Fox}\email[Corresponding author: ]{pjfox@fnal.gov}
\author{Sergo Jindariani}
\author{Vladimir Shiltsev}
\affiliation{%
\\
\mbox{Fermi National Accelerator Laboratory, Batavia Illinois 60510-5011, USA}\\
}%
\date{\today}
\preprint{FERMILAB-CONF-22-010-AD-PPD-T}

\begin{abstract}
\snowmass{}
\section*{Executive Summary}
While dimuonium $(\mu^+\mu^-)$ -- the ``smallest QED atom" -- has not yet been observed, it is of utmost fundamental interest.  By virtue of the larger mass, dimuonium has greater sensitivity to beyond the standard model (BSM) effects than its cousins positronium or muonium, both discovered long ago, while not suffering from large QCD uncertainties.
Dimuonium atoms can be created in $e^+e^-$ collisions with large longitudinal momentum, allowing them to decay a small distance away from the beam crossing point and avoid prompt backgrounds. We envision a unique cost-effective and fast-timeline opportunity for copious production of $(\mu^+\mu^-)$ atoms at the production threshold via a modest modification of Fermilab's existing FAST/NML facility to arrange collisions of 408 MeV electrons and positrons at a 75$^{\rm o}$ angle. This compact 23 m circumference collider (DIMUS) will allow for precision tests of QED and open the door for searches for new physics coupled to the muon.  Fermilab's FAST/NML is perfectly suited for DIMUS as there are existing SRF accelerators and infrastructure, capable of producing high energy, high current electron and positron beams, sufficient for $O(10^{32})\mathrm{cm}^2\mathrm{s}^{-1}$ luminosity and $\sim$0.5 million dimuons per year. The expansion will require installation of a second SRF cryomodule, positron production and accumulation system, fast injection/extraction kickers and two small circumference intersecting rings. An approximately meter-sized detector with several layers of modern pixelated silicon detector and crystal-based electromagnetic calorimeters will ensure observation of the decays of dimuonium to electron-positron pairs in presence of the Bhabba  scattering  background. An expansion of the system to include  solenoidal magnet outside of the calorimeter system, a layer of steel shielding behind the magnet, and a set of dedicated muon detectors would extend the physics program of DIMUS to include precision studies of rare processes with muons, pions, and $\eta$ mesons produced in $e^{+}e^{-}$ collisions. \end{abstract}
\maketitle
\section{Introduction}
We discuss a low energy $e^+e^{-}$  collider for production of the not yet observed $\mu^+\mu^-$ bound system, often called Dimuonium or True Muonium (TM), at the threshold $\sqrt{s}=211$ MeV - Dimuonium Spectroscopy collider (DIMUS) - on the Fermilab site. This relatively small collider with electron and positron orbits intersecting at a 75$^{\rm o}$ crossing angle produces dimuonium with non-zero forward momentum.  This shifts its decay point out of the beam collision area and provides effective suppression of the elastic $e^+e^-$ scattering background. We present preliminary considerations of the required accelerator complex (full energy $e^+$ and $e^{-}$  injectors and 23 m circumference collider ring) based on existing high-flux electron SRF accelerator complex at the NML (FAST facility) at Fermilab and discuss its main parameters, as well as discuss the requirements for efficient detection of the produced dimuonium. The main preliminary parameters of the DIMUS at NML/FAST are listed in Table \ref{T1}.  High luminosity $\mathcal{O}(10^{32})$cm$^{-2}$s$^{-1}$ allows detailed studies of the dimuonium system and may open a new window in the search for beyond the standard model physics. 

\begin{table}[h]
\begin{tabular}{|lcc|}
\hline 
\hline
Beam energy & 408 & MeV \\
C.m.e. $\sqrt{s}$ & 211 & MeV \\
C.m.e. spread & 0.4 & MeV \\
Crossing angle & 75 & deg. \\
Circumference & 23 & m \\
Beta-functions at IP ($y, x$) & 20/0.2 & cm \\
Bunch length & 1.2 & cm \\
Bunch spacing & 1.9 & ns \\
Beam sizes at IP ($y,x$) & 0.7/130 & $\mu$m \\
Number of bunches & 40 & \\
Number of $e^+/e^-$ per bunch & 4$\cdot$10$^{10}$ & \\
Beam lifetime & $\ge 30$ & sec \\
Max $e^+$ production rate & 4$\cdot$ 10$^{10}$ & $e^+$/s \\
Peak luminosity & 1.6 $\cdot$ 10$^{32}$ & cm$^{-2}$s$^{-1}$\\
\hline \hline
\end{tabular}
\caption{Main parameters of the DIMUS collider at Fermilab's NML/FAST.} 
\label{T1}
\end{table}

A study of the potential of the LHCb experiment to discover, for the first time, the true muonium bound state was reported in Ref.~\cite{CidVidal:2019qub}. The study suggests that an observation significance can be achieved with approximately 15 $fb^{-1}$ of LHC Run-3 data.
However, it relies on the $\eta\rightarrow\gamma+TM$ process and assumes that a soft photon originating in the $\eta$ meson decay can be efficiently reconstructed, which may turn out to be challenging. A possibility of observing true muonium at running (HPS, DIRAC) and proposed (REDTOP) fixed-target experiments has also been discussed in the literature~\cite{Lamm:2017ioi}. Finally, a dedicated electron-positron collider was proposed for construction in Novosibirsk and is described in Ref.~\cite{Bogomyagkov:2017uul}. To date, no measurements with data have been reported. 

\section{Theory}

While electromagnetically bound states are ubiquitous in nature, so far the ``smallest QED atom" \cite{Brodsky:2009gx} $(\mu^+\mu^-)$ has not yet been observed.  By virtue of the larger reduced mass of the system ($m_\mu/2$) dimuonium has larger sensitivity to beyond the standard model (BSM) effects than its cousins positronium or muonium, both discovered over 60 years ago \cite{Deutsch:1951zza,Hughes:1960zz}, while not suffering from large QCD uncertainties.  

\begin{figure}[h]
\centering
\includegraphics[width=0.8\linewidth]{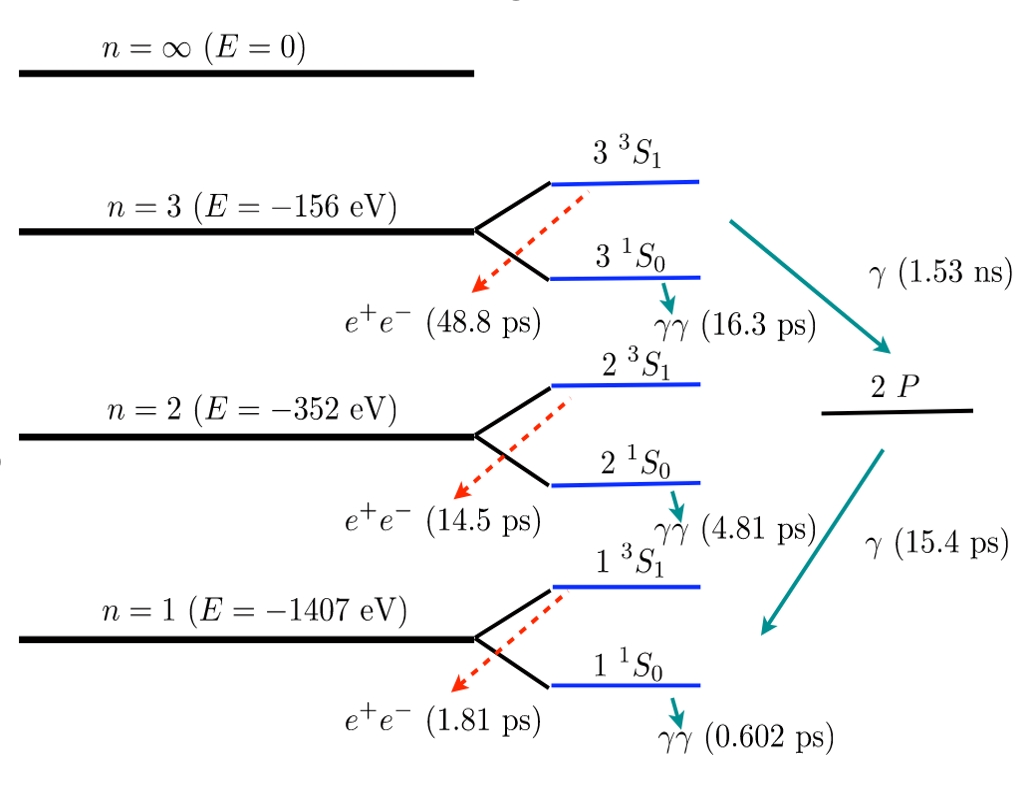}
\caption{Schematic of the energy levels of the dimuon system along with and their decays and transition times, taken from \cite{Brodsky:2009gx}.} 
\label{fig:spectrum}
\end{figure}

The spectrum of the dimuon system can be found by rescaling that of positronium, with the lowest bound state appearing $\sim 1.4$ keV below the two muon threshold.  The QED corrections to the leading order energy levels are known to $\mathcal{O}(\alpha^5)$.  The decays of the bound states require overlap of the muon wavefunctions and thus only occur in $\ell=0$ states.  In addition to the positronium-like decay to two photons dimuonium has an additional available decay mode to an electron-positron pair.  Thus, the decay lifetimes of the ortho and para states are more similar than for positronium, where ortho-positronium must decay to three photons.  In Fig.~\ref{fig:spectrum} we show a schematic of the energy levels and their decays/transitions.  The energy levels are denoted as $n^{2s+1}\ell_J$ and the lifetimes are given by
\begin{equation}
\tau(n^3 S_1\rightarrow e^+ e^-)=3\tau(n^3 S_0\rightarrow \gamma\gamma)=\frac{6n^3}{\alpha^5 m_\mu}~,
\end{equation}
note that these decay times are appreciably shorter than the muon lifetime and the muon can be treated as a stable particle.

The DIMUS collider produces dimuons in any of the $n^3 S_1$ states, with relative rate scaling as $n^{-3}$.  Since the beams cross with intersecting angle $\theta\approx 75^\circ$ the bound states are produced with a boost, $\gamma=\tan \theta$, which is critical to allow the dimuon system to move away from the beam crossing region (and its associated prompt backgrounds) before decaying.  The production cross section is slightly Sommerfeld-enhanced from the naive expectation due to the small relative motion, $v_{\mathrm{rel.}}\sim \alpha$, of the muons.  For electron/positron beams with an energy spread $\Delta E_e$ the ratio of bound state production cross section to the Bhabha scattering background is $\sigma_{\mathrm{b.s.}}/\sigma_{\mathrm{Bhabha}}\sim 10^{-4} (1\,\mathrm{MeV}/\Delta E_e)$.  The expected number of dimuonia produced per year is $N_{(\mu^+\mu^-)}\sim 10^5 (1\,\mathrm{MeV}/\Delta E_e) (\mathcal{L}_{\mathrm{int}}/10^{39} \mathrm{cm}^2)$.

Once a dimuonium state is produced it can be studied in detail, allowing for a precision test of QED and a probe of BSM physics.  Although produced only in $S_1$ states it may be possible to populate a $P$ state by passing the dimuonium through a strong magnetic field or microwave laser \cite{Brodsky:2009gx} allowing measurement of the Lamb shift and other precision spectroscopic measurements.  Furthermore, by measuring decay times, production rates, and searching for new decay modes bounds (or a discovery) of BSM physics can be made.  The ongoing anomalies in the muon magnetic moment \cite{PhysRevLett.126.141801}, the proton radius puzzle (see \cite{Paz:2019wfq} and references therein), and the evidence of breaking of lepton universality in $B$ decays \cite{LHCb:2021trn} may point to BSM physics coupled to muons and motivate searches through dimuonium.  This new muonic force will distort the muon wavefunctions in dimuonium and can alter production and decay, or the energy levels and thus could reveal itself in spectroscopic measurements.  Given the existing constraints on muonic forces these BSM corrections should not be larger than $\sim 100$ MHz, which would require additional orders of precision on the theory side, however there is no obvious impediment to these calculations \cite{Lamm:2015fia}.  If the new physics contains a light boson $X$, then new decay modes such as $(\mu^+\mu^-)\rightarrow \gamma X, XX$ may be open, and $X$ may further decay to a pair of SM states (e.g. photons or electrons) or to BSM states (e.g. dark matter).  Unlike indirect observables, like muon $(g-2)$, observation of such a decay mode can distinguish between different model scenarios \cite{Tucker-Smith:2010wdq,CidVidal:2019qub}.

\section{Accelerator}
\begin{figure}[htbp]
\centering
\includegraphics[width=0.8\linewidth]{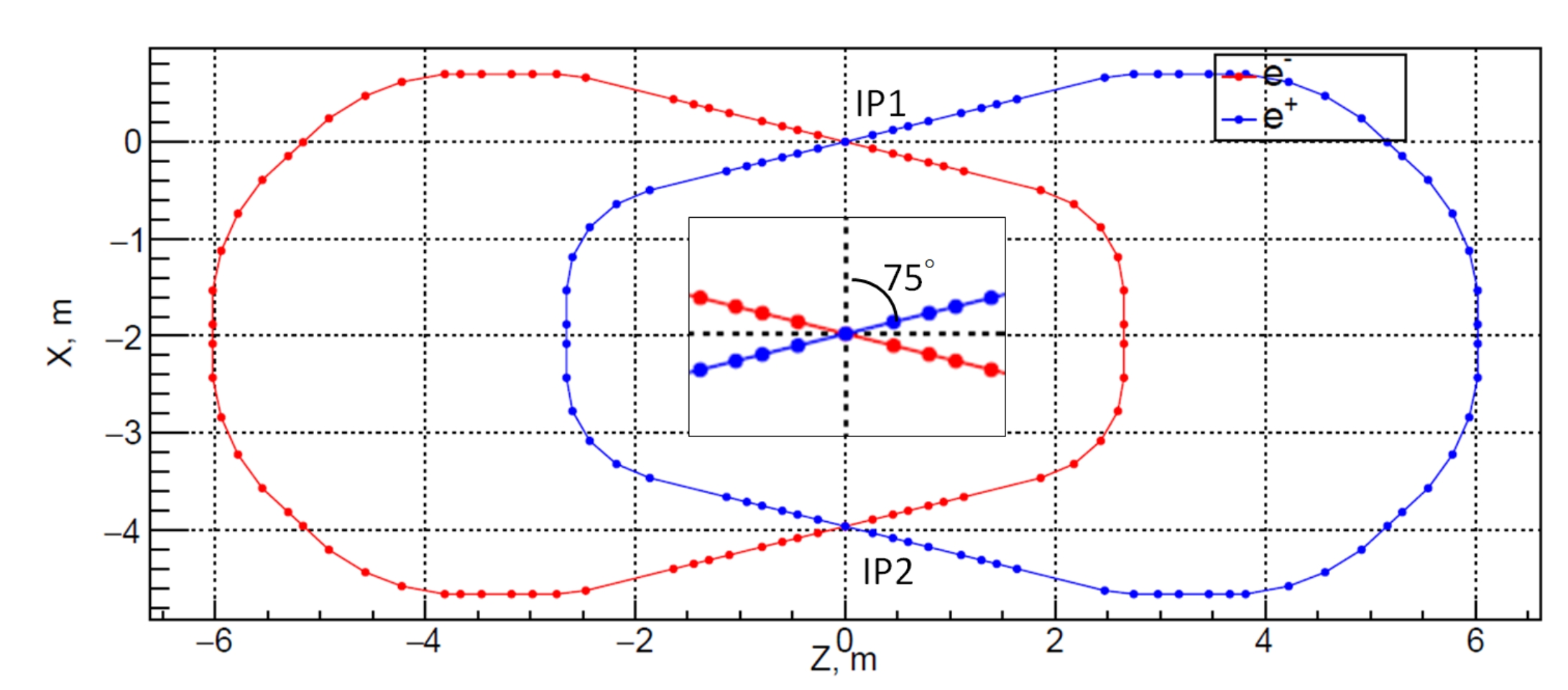}
\caption{Schematics of the intersecting storage rings.} 
\label{fig2}
\end{figure}

Effective accelerator lattice design of the electron-positron collider for $\mu^+\mu^-$ search and studies has been discussed in \cite{bogomyagkov2018low}. Low beam energy $E_b=408$ MeV results in a compact (12 m$\times$6 m footprint), simple and symmetric configuration collider with a large 75$^{\rm o}$ crossing angle of the intersecting beams to arrange a non-zero dimuonium longitudinal momentum. Schematics of the two intersecting bulb-shape storage rings with two intersection points is shown in Fig.~\ref{fig2}.
The main challenges of such collider design include: a) the need in high positron production rate $O(10^{11} s^{-1})$, required for high luminosity; b) low beam energy and high bunch intensity result in strong intra-beam scattering (IBS) and reduced Touschek lifetime, c) large collision angle leads to a number of potentially damaging effects in $e^+e^-$ beams interaction. 
The demand for the positron production is set by the total number of positrons circulating in the collider and the beam lifetime which in turn depends on an interplay of beam beam effects, intrabeam scattering and Touschek effect - see Table \ref{T1}. The existing high-flux electron SRF accelerator complex at the NML (FAST facility) at Fermilab is an excellent source of high energy electrons in the energy range of 100-300 MeV \cite{Broemmelsiek:2018iqr}. Its maximum operational $e^-$ production capacity is 3000 bunches$\times$5 Hz$\times 2\cdot 10^{10}=3\cdot 10^{14} e^-$/s \cite{Broemmelsiek:2018iqr}. In order to have full energy injection into the collider at 408 MeV, another 1.3 GHz SRF cryomodule will need to be installed - and the NML/FAST facility has plenty of space and all the required infrastructure. Note, that the demand of 4$\cdot$ 10$^{10}\ e^+$/s for DIMUS does not require full operational capacity of the SRF linac - it can be obtained with 1\% $e^- \rightarrow e^+$ conversion efficiency with a mere 200 bunches$\times$1 Hz$\times 2\cdot 10^{10} = 4\cdot 10^{12} e^-$/s. 

\begin{figure}[htbp]
\centering
\includegraphics[width=0.8\linewidth]{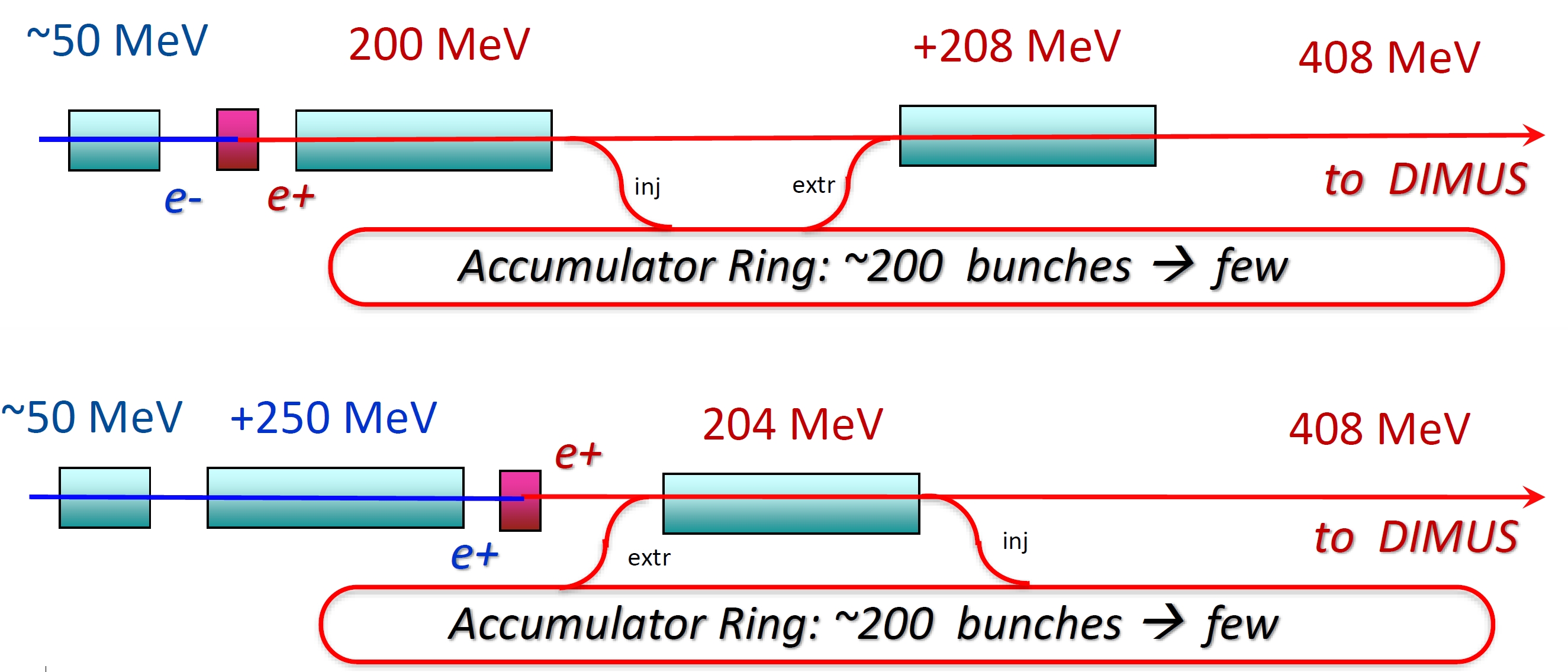}
\caption{Schematics of two possible schemes of positron production, damping and injection for DIMUS.} 
\label{fig3}
\end{figure}

The positron production complex - see Fig.~\ref{fig3} - does not exist now and will need to include i) conversion of either 50 MeV or 300 MeV electrons (pulse of 200 bunches 333 ns apart) to positrons and a collection system; ii) acceleration of positrons to either $\sim$ 200 MeV and their bunch-by-bunch injection into a 120 m long accumulator (damping) ring where they are spaced by 2 ns; after sub-second damping time 200 low intensity positron bunches collapse into one with $\sim$4$\cdot 10^{10}$  $e^+$'s; ii) then the single positron bunch is extract and accelerated to 408 MeV before being injected into the 23 m long ($\sim$80 ns) DIMUS $e^+$ ring, to become one of 40  colliding $e^+$  bunches. 

The above injection/extraction manipulations will require fast kickers with 4 ns EM pulse duration, similar to ones developed and tested for TESLA and ILC colliders \cite{Grishanov:1996wr, Naito:2011zz}.

\section{Detector}
Design of the detector for the DIMUS collider is strongly coupled to the physics goals of such a machine. Here we will focus on the $e^+e^-$ decay of the true muonium. While other decays modes are possible, they are more challenging in terms of detection and reconstruction. The primary physics background for the $e^+e^-$ mode will consist of Bhabba scattering events with production rates three orders of magnitude larger than those of the signal. However, a number of handles are available to suppress the background. Firstly, due to their non-zero lifetime, true muonium states will travel approximately 2 mm before decaying into $e^+e^-$, while electrons and positrons in Bhabba scattering process will originate at the interaction region. It is therefore imperative to achieve excellent spacial resolution for the $e^+e^-$ vertex reconstruction.  

\begin{figure}[htbp]
\centering
\includegraphics[width=0.8\linewidth]{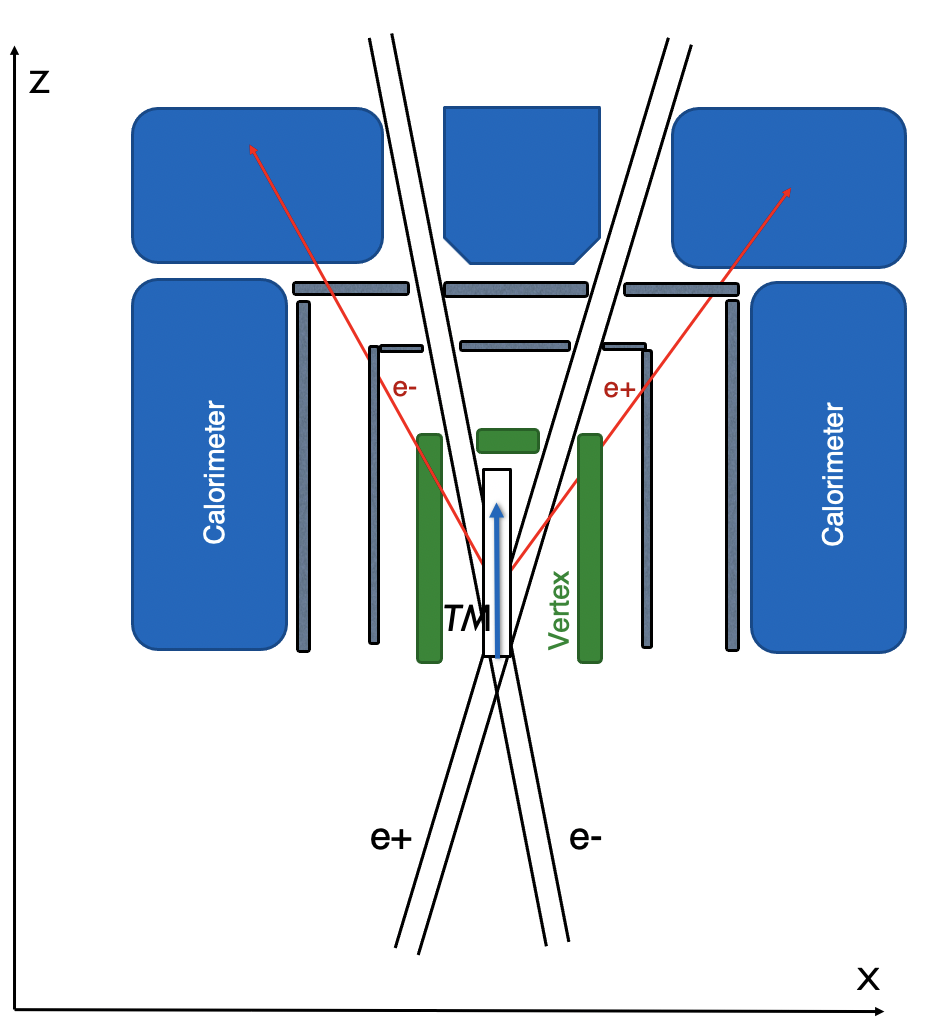}
\caption{Schematic of a detector for the DIMUS collider. Silicon pixel detector is shown in green with electromagnetic calorimeters illustrated by the blue boxes. Two silicon or gas detector layers shown in grey provide extra tracking capabilities for precise directionality and linking of electron/positron tracks to the EM clusters.} 
\label{fig5}
\end{figure}

The size of the interaction region is expected to be approximately 300 microns along the direction of the beam $z$. In order to contribute negligibly to the position resolution, a detector capable of reconstructing $e^+e^-$ vertex with precision of better than 100 $\mu m$ along the beam is needed. Pixelated silicon detectors with 3-4 layers of sensors and 100-150 $\mu m$ pixel size should be able to provide the target resolution. With the combined vertex $z$-position resolution of better than 400 $\mu m$, a cut of $v_{z}>2$ mm would allow the reduction of the Bhabba background to a very small level. 

Furthermore, one can explore the resonant nature of the electron-positron invariant mass $M(e^+e^-)$ for the signal, while discriminating it against the continuous Bhabba background. The typical energy of electrons and positrons originating in true muonium decays is of the order of a hundred MeV, while transition between the true muonium states happens with the emission of photons in the 100 eV -- 2 KeV range. It is therefore desirable to achieve electron/positron energy resolution at the $10^{-2}$ level in the electromagnetic calorimeter and even better $10^{-3}-10^{-4}$ for any spectroscopy measurements. Since there will be only few particles arriving into the detector per beam crossing, granularity is a secondary consideration. Crystal based calorimetery provides a natural solution to the problem, with LYSO crystals producing very high light yields and therefore achievable energy resolutions of $\mathcal{O}(1)\%$ or better.  Other crystals, such as PbW04 can also be considered but are not as bright as LYSO and may not have enough light to achieve the target energy resolution. 

An additional 2-3 tracking layers made of silicon strips or gas detectors (e.g. GEM) will provide a link between hits reconstructed in the pixel detector and the calorimeter clusters and ensure high precision determination of the electron and positron direction. A schematic of a detector for the DIMUS collider based on the presented considerations is shown in Fig.~\ref{fig5} with a rough size of about one meter. It is expected that a detector like this could achieve better than 50\% acceptance per track, leading to 25+\% total acceptance to signal events. This would allow for 100,000-500,000 of reconstructable signal events per year.

As mentioned earlier, spectroscopy studies of the true muonium states would require electron energy resolution of $10^{-3}-10^{-4}$ which is extremely difficult to achieve. Another possibility is to detect and measure energy of the x-ray photons produced in the transitions. Detection of such photons has been studied in detail at the kaon-nucleus  experiments using different technologies. These include Silicon Lithium (SiLi), charge-coupled devices (CCD), silicon drift detectors (SDD). SDD devices are particularly promising due to a combination of relatively large sensor area, good timing capabilities and excellent energy resolution. However, feasibility of their application for a DIMUS experiment needs to be carefully evaluated taking into account the system size, incident particle rates, and cost considerations. The SDDs will need to be placed close to the interaction region. Some shielding will likely be necessary in order to reduce the exposure of the SDDs to beam backgrounds. This will lead to a reduced detection acceptance for the x-ray photons for the electrons produced in the dimuonium decay. An alternative approach based on the laser spectroscopy has been previously mentioned in the literature but would have to be carefully evaluated taking into account the complexity and power need for such an approach.

Finally, it should be mentioned that a more complex detector would be necessary in order to extend the physics program of DIMUS to include precision studies of rare processes with muons, pions, and $\eta$ mesons produced in $e^{+}e^{-}$ collisions. This includes a solenoidal magnet outside of the calorimeter system, a layer of steel shielding behind the magnet, and a set of dedicated muon detectors. The size of such a detector will be in the several meter range. 

\section{Conclusion}
Dimuonium atoms are of fundamental interest, and so far have not been observed. They can be created in $e^+e^-$ collisions with large longitudinal momentum, allowing them to decay a small distance away from the beam crossing point and avoid prompt backgrounds e.g. 408 MeV/beam at 75$^{\rm o}$. Fermilab's FAST/NML is perfectly suited for DIMUS as there are existing SRF accelerators and infrastructure, capable of producing copious amounts of electrons and positrons, sufficient for $O(10^{32})\mathrm{cm}^2\mathrm{s}^{-1}$ luminosity and $\sim$0.5 million dimuons per year. 
The facility expansion to the collider will require installation of the second SRF cryomodule, positron production and accumulation system, and two small circumference collider rings.  Furthermore, observation of the decays of dimuonium to electron-positron pairs will require building a meter-sized tracking detector.  The copious production of dimuonium at DIMUS allows for precision tests of QED and opens the door for searches for new physics coupled to the muon. 

\section{Acknowledgements}
Work supported by the Fermi National Accelerator Laboratory, managed and operated by Fermi Research Alliance, LLC under Contract No. DE-AC02-07CH11359 with the U.S. Department of Energy. The U.S. Government retains and the publisher, by accepting the article for publication, acknowledges that the U.S. Government retains a non-exclusive, paid-up, irrevocable, world-wide license to publish or reproduce the published form of this manuscript, or allow others to do so, for U.S. Government purposes.

\bibliography{references}

\end{document}